\documentclass[twocolumn,prl,showpacs,preprintnumbers,amsmath,amssymb]{revtex4}

\usepackage[latin1]{inputenc}
\usepackage[OT1]{fontenc}
\usepackage{graphicx}
\usepackage{dcolumn}
\usepackage{bm}

\begin{document}

\preprint{APS/123-QED}

\title{Delayed-choice test of complementarity with single photons}

\author{Vincent Jacques$^1$, E Wu$^{1,2}$, Fr\'ed\'eric Grosshans$^1$, Fran\c{c}ois Treussart$^1$, Philippe Grangier$^3$, Alain Aspect$^3$ and Jean-Fran\c{c}ois Roch$^1$}

\affiliation{ $^1$Laboratoire de Photonique Quantique et Mol\'eculaire, Ecole Normale Sup\'erieure de Cachan, UMR CNRS 8537, Cachan, France\\
$^2$State Key Laboratory of Precision Spectroscopy, East China Normal University, Shanghai, China\\
$^3$Laboratoire Charles Fabry de l'Institut d'Optique, UMR CNRS 8501, Palaiseeau, France}
\date{\today}

\begin{abstract}
We report an experimental test of complementarity using clock-triggered single-photon pulses emitted by an individual N-V color center in a diamond nanocrystal. The single photons are sent into a Mach-Zehnder interferometer with an output beamsplitter of adjustable reflection coefficient $R$. In addition, the choice of introducing or removing this beamsplitter is random and relativistically space-like separated from the entering of the photon inside the interferometer, as required for the Wheeler's delayed-choice regime.  Each set value of $R$ allows us to observe interference with \textit{visibility} $V$ and to obtain incomplete which-path information characterized by the \textit{distinguishability} $D$. The measured values of $V$ and $D$ are found to obey the complementarity relation $V^{2}+D^{2}\leq 1$.
\end{abstract}

\pacs{Valid PACS appear here}
\maketitle

As emphasized by Bohr~\cite{Bohr}, complementarity lies at the heart of quantum mechanics. A celebrated example is the illustration of wave-particle duality by considering single particles in a two-path interferometer~\cite{Feynman}, where one chooses either to observe interference fringes, obviously associated to a wave-like behavior, or to know which path of the interferometer has been followed, according to a particle-like behavior~\cite{footnote no entanglement}. Although interference has been observed at the individual particle level with electrons \cite{Tonomura}, neutrons \cite{Neutrons}, atoms~\cite{Carnal,Keith}, molecules~\cite{Arndt}, only a few experiments with massive particles have explicitly checked the mutual exclusiveness of which-path information (WPI) and interference~\cite{Pfau,Pritchard,Buks,RempeNature,Haroche}. \\
\indent In the case of photons, it has been pinpointed that meaningful two-path interference experiments demand a single-photon source~\cite{Grangier} for which full and unambiguous WPI can be obtained, complementary to the observation of interference~\cite{Grangier,Kurtsiefer,Jacques}.  In order to rule out a too naive view of complementarity, which would assume that the particle could ``know'' when entering the apparatus which experimental configuration has been set (record of interference or determination of WPI) and would then adjust its behavior accordingly~\cite{Greenstein}, Wheeler proposed the ``delayed choice" scheme where the choice between the two complementary measurements is made long after the particle entered the interferometer \cite{Wheeler}. Realizations of that gedanken experiment \cite{Zajonc,Martiennsen,Baudon,DCECachan} have confirmed that the chosen observable can be determined with perfect accuracy even if the choice, made by a quantum random number generator, is space-like separated from the entering of the particle into the interferometer \cite{DCECachan}.\\
\indent In 1978, Wooters and Zurek~\cite{Zurek} considered an intermediate situation in which interaction with the interfe\-rometer considered as a quantum device allows one to gain an imperfect -- but significant -- knowledge of WPI, without destroying the interference pattern, which remains observable with a good -- although reduced -- visibility. In 1988, Greenberger and Yasin noticed that in an unbalanced interferometer as used in some neutron interferometry experiments, one has partial WPI while keeping interference with limited visibility~\cite{GreenbergerYasin}. The complementary quantities WPI and interference visibility could then be partially determined simultaneously.\\
\indent Consistent theoretical analysis of both schemes, independently published by Jaeger et al.~\cite{Jaeger_PRA1995} and by Englert~\cite{Englert} leads to the inequality~\cite{inequality} 
\begin{equation}
\label{ComplIneq}
V^{2}+D^{2}\leq 1
\end{equation}
 which puts an upper bound to the maximum values of simultaneously determined interference visibility $V$ and path distinguishability $D$, a parameter that quantifies the available WPI on the quantum system.\\
\indent The all-or-nothing cases $(V=1,D=0)$ or $(V=0,D=1)$~\cite{Tonomura,Neutrons,Carnal,Keith,Arndt,Grangier,Kurtsiefer,Jacques} obviously fulfill inequality~\eqref{ComplIneq}. Intermediate situations, corresponding to partial WPI and reduced visibility, have been investigated using atoms \cite{Rempe}, nuclear spins \cite{Peng} and faint laser light~\cite{attenuated light pulses}. However none of them has been realized in the delayed-choice scheme. We report here an experimental test of the complementarity inequality~\eqref{ComplIneq} in intermediate regimes using true single-photon pulses, and in the delayed-choice operation mode.\\
\indent Following Englert~\cite{Englert}, we point out that the distinguishablility $D$ constrained by inequality~\eqref{ComplIneq} actually corresponds to two different notions. The \textit{a priori} distinguishability, also called ``predictability", refers to a WPI obtained by using an unbalanced interferometer with different particle flux along the two paths. Only the case where path distinguishability is introduced \textit{a posteriori}, \textit{i.e.} after the entering of the particles into the interferometer, offers the opportunity of a delayed choice test of complementarity. This \textit{a posteriori} distinguishability can be introduced either by creating entanglement between the particle and a which-path marker~\cite{Haroche,ScullyNature} or by using an interferometer with an unbalanced output beamsplitter~\cite{Rempe}. We have chosen the latter case by implementing  the scheme depicted on Fig.~1, where a single-photon pulse is sent into a Mach-Zehnder interferometer with a variable output beamsplitter (VBS) of adjustable reflection coefficient $R$. When $R$ is not $0.5$, one can have some WPI by observing which detector ($\rm P_{1}$ or $\rm P_{2}$) is fired. The choice of introducing or removing this beamsplitter is random and relativistically space-like separated from the entering of the photon inside the interferometer, as required for the delayed-choice regime.\\
\begin{figure}[t]
    \centerline{\includegraphics[width=8cm]{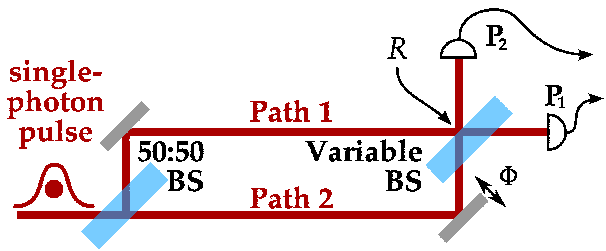}}
    \caption{Theroretical sketch of the delayed-choice complementarity-test experiment. A single-photon pulse is sent into a Mach-Zehnder interferometer, composed of a 50/50 input beamsplitter (BS) and a variable output  beamsplitter (VBS). The reflection coefficient is randomly set either to the null value or to an adjustable value $R$, after the photon has entered the interferometer. The single-photon photodetectors $\rm P_{1}$ and $\rm P_{2}$ allow to record both the interference and the WPI. 
 }
        \label{FigSetup}
        
    \end{figure}
\indent The experiment starts from a clock-triggered single-photon source, based on the photoluminescence of a single N-V color center in a diamond nanocrystal~\cite{Kun}. The linearly polarized single-photon pulses are then directed to a polarization Mach-Zehnder interferometer described in Ref.~\cite{DCECachan}. The input polarization beamsplitter $\rm BS$ splits the light pulse into two spatially separated components of equal amplitudes, associated with orthogonal S and P polarizations. The two beams then propagate in free space for 48~m, which corresponds to a  time of flight of 160~ns.\\
\indent The variable output beamsplitter VBS is the association of a polarization beamsplitter (PBS) which spatially overlaps the two beams, an electro-optical modulator (EOM)  which acts as an adjustable waveplate, and a Wollaston prism (WP) with its polarization eigenstates corresponding to the S and P polarized channels of the interferometer (Fig.~\ref{FigVBS}). Given the relative orientation $\beta$ of the EOM, the VBS reflection coefficient $R$   depends on the voltage  $V_{\rm EOM}$ applied to the EOM, according to the relation:
\begin{equation}
R=\sin^{2} 2\beta \times \sin^{2}\left(\frac{\pi}{2} \frac{ V_{\rm EOM}}{  V_{\pi}}\right)
\label{defR}
\end{equation}
where $V_{\pi}$ is the half-wave voltage of the EOM. The parameters $\beta$ and $V_{\pi}$ have been independently measured for our experimental conditions, and found equal to $\beta=24\pm 1°$ and $V_{\pi}=217\pm 1\ \rm V$ at the wavelengh $\lambda=670 \ \rm nm$ which is the emission peak of the negatively charged N-V color center~\cite{Kun}. This allows $R$ to vary between $0$ and $0.5$ when $V_{\rm EOM}$ is varied between $0$ and $170$ V.\\
\indent When $R=0$, the VBS is equivalent to a perfectly transparent (or absent) beamsplitter. Then, each ``click'' of one of the two photodetectors ($\rm P_{1}$ or $\rm P_{2}$) placed on the output ports of the interferometer is associated to a specific path. It then gives access to the full WPI ($D=1$) and no interference effect will be observed ($V=0$). When $R\neq0$, paths $1$ and $2$ are partially recombined by the VBS. The WPI is then partially washed out, up to be totally erased when $R=0.5$. On the other hand, interference can be observed when the dephasing $\Phi$ between paths 1 and 2 is varied. The experiment will consist in checking the relation between $D$ and $V$ for a given value of $R$, controlled by the EOM voltage $V_{\rm EOM}$.

\begin{figure}[b]
    \centerline{\includegraphics[width=8cm]{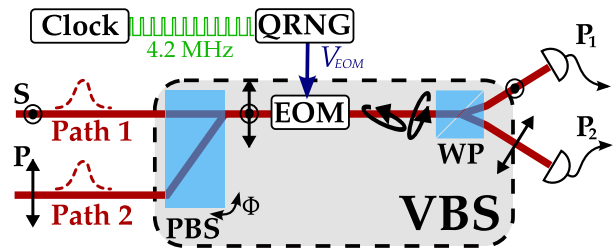}}
    \caption{Variable output beamspltter (VBS) implementation. The optical axis of the polarization beamsplitter (PBS) and the polarization eigenstates of the Wollaston prism (WP) are aligned, and make an angle $\beta$ with the optical axis of the EOM. The voltage $V_{\rm EOM}$ applied to the EOM is randomly chosen accordingly to the output of a Quantum Random Number Generator (QRNG), located at the output of the interferometer and synchronized on the 4.2~MHz clock  that  triggers the single-photon emission. }
        \label{FigVBS}
 \end{figure}
 
\indent In order to perform the experimental test of complementarity in the delayed-choice regime, the choosen configuration of the interferometer, defined by $R$, has to be causally isolated from the entering of the photon into the interferometer. This condition is ensured by a relativistically space-like separated random choice. For each measurement, the value of the reflection coefficient of VBS is randomly chosen between $0$ and a given value of $R$, using a quantum random number generator (QRNG) located at the output of the interferometer (Fig. 2)~\cite{DCECachan}. The random numbers are generated from the amplified shotnoise of a white light beam which is an intrinsic quantum random process. At the experiment clock-frequency, \textit{i.e.} every $\tau_{\rm rep}=238$ ns, fast comparison of the amplified shotnoise to the zero level generates a binary random number $0$ or $1$ which changes the VBS reflectivity between $0$ and $R$, by applying or not the corresponding voltage to the EOM (see Eq.~\eqref{defR}). In the laboratory framework, the random choise is realized simultaneously with the entering of the photon into the interferometer, ensuring the required space-like separation~\cite{DCECachan}. \\
\indent As meaningfull illustration of complementarity requires the use of single particles, the quantum behavior of the light field is first tested using the two output detectors feeding single and coincidence counters with no voltage applied to the EOM. In this situation of an absent output beamsplitter, we measure the correlation parameter $\alpha$~\cite{Grangier,Jacques}, which is equivalent to the second order correlation function at zero delay $g^{(2)}(0)$. For an ideal single-photon source, quantum optics predicts a perfect anticorrelation $\alpha=0$, in agreement with the particle-like image that the photon cannot be detected simultaneously in the two paths of the interferometer. With our source~\cite{Kun}, we find $\alpha = 0.15 \pm 0.01$, a value much smaller than one, showing that we are indeed close to the pure single-photon regime~\cite{alpha}.
 \begin{figure}[t]
\includegraphics[width=8cm]{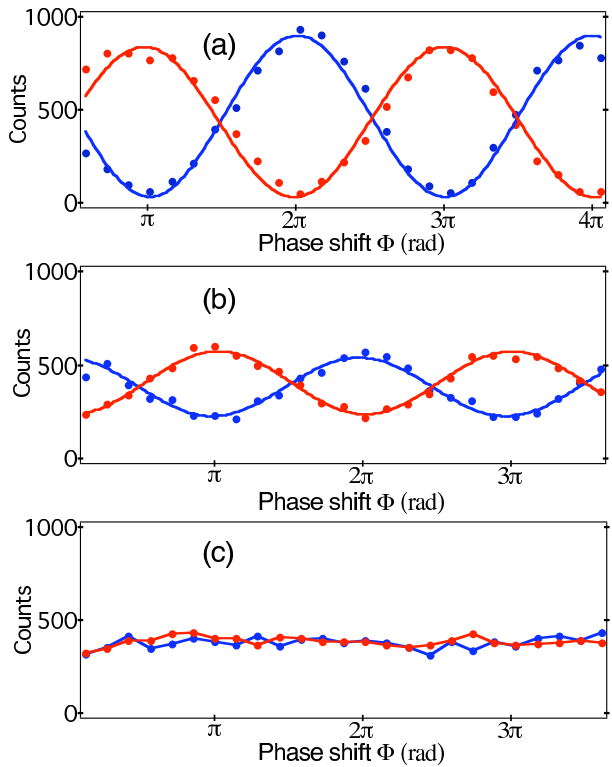}
\caption{Interference visibility $V$ measured in the delayed choice regime for different value of $V_{\rm EOM}$ applied randomly to the EOM. (a), (b), (c) correspond respectively to $V_{\rm EOM}\approx 150 \ $V ($R=0.43$ and $V=93\pm 2\%$), $V_{\rm EOM}\approx 40$ \ V ($R=0.05$ and $V=42\pm 2\%$) and $V_{\rm EOM}=0$ ($R=0$ and $V=0$). Each point is recorded with $1.9$ s acquisition time. Detectors dark counts, of about $60 \ \rm s^{-1}$ for each, have been substracted to the data.}
\label{Figresult}
\end{figure}

\indent The delayed-choice test of complementarity with single-photon pulses is performed with the EOM randomly switched for each photon sent in the interferometer, corresponding to a random choice between two values $0$ and $R$ of the VBS reflectivity. The phase-shift $\Phi$ between the two arms of the interferometer is varied by tilting the polarization beamsplitter PBS of VBS with a piezoelectric actuator (see Fig.\ref{FigVBS}). For each photon, we record the chosen configuration of the interferometer, the detection events, and the actuator position. All raw data are saved in real time and are processed only after a run is completed. The events corresponding to each configuration of the interferometer are finally sorted. For a given value $R$, the wave-like information of the light field is obtained by measuring the visibility of the interference, predicted to be
  \begin{equation}
    V=2\sqrt{R(1-R)} \ .
    \label{defV}
    \end{equation}
    \begin{figure}[b]
\includegraphics[width=8cm]{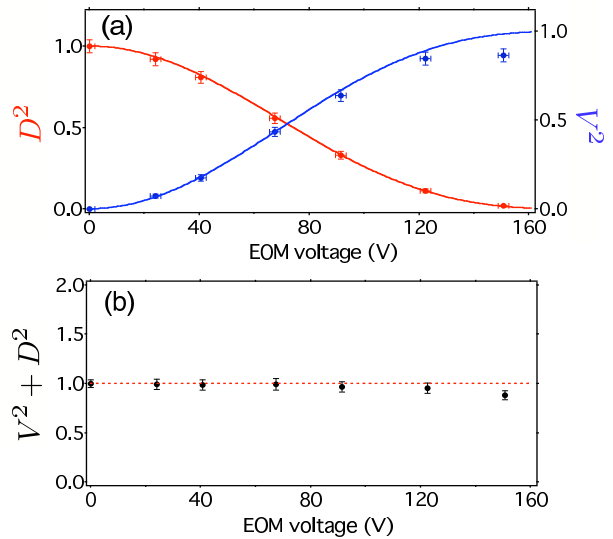}
\caption{\label{Figresult}
Delayed-choice test of complementarity with single-photon pulses. (a)-Wave-like information $V^{2}$ and which-path information $D^{2}$ as a function of the EOM voltage corresponding to a given value $R$ of the VBS reflectivity. The solid lines are the theoretical expectations, with $\beta=24 °$ and $V_{\pi}=217$, using Eqs~\eqref{defR},~\eqref{defV} and~\eqref{Predic}. (b)- $V^{2}+D^{2}$ as a function of the EOM voltage.}
\end{figure}
\indent The results, depicted in Fig. 3, show a reduction of $V$ when the value of $R$ randomly applied decreases.\\
\indent To test inequality \eqref{ComplIneq}, a value of the distinguishability $D$ is then required, to qualitatively qualify the amount of WPI which can be extracted for each value of $R$. We introduce the quantity $D_{1}$ and $D_{2}$, respectively associated to the WPI on path 1 and path 2: 
\begin{align}
D_{1}=&\left|p({\rm P}_{1},{\rm path} \ 1)-p(\rm P_{2},\rm path \ 1)\right|\\
D_{2}=&\left|p({\rm P}_{1},{\rm path} \ 2)-p(\rm P_{2},\rm path \ 2)\right|
\label{defI1}
\end{align}
where $p(\rm P_{i},\rm path \ j)$ is the probability that the particle follows path j and is detected on detector $\rm P_{i}$. For a single particle arriving on the output beamsplitter, one obtains 
\begin{equation}
D_{1}=D_{2}=\frac{1}{2}-R \ .
\end{equation}
The distinguishability parameter $D$ is finally defined as~\cite{Englert}
\begin{equation}
D=D_{1}+D_{2}=1-2 R \ .
\label{Predic}
\end{equation}
In order to test this relation, we estimate the values of $D_{1}$ and $D_{2}$ by blocking one path of the interferometer and measuring the number of detections $N_{1}$ and $N_{2}$ on detectors $\rm P_{1}$ and $\rm P_{2}$, which are statisticallly related to $D_{1}$ and $D_{2}$ according to~\cite{Rempe,DCECachan} :
\begin{align}
D_{1}=&\left.\left|\frac{N_{1}-N_{2}}{N_{1}+N_{2}} \right| \right]_{\text{path 2 blocked}}\\
D_{2}=&\left.\left|\frac{N_{1}-N_{2}}{N_{1}+N_{2}} \right| \right]_{\text{path 1 blocked}}.
 \end{align}
 \indent These measurements are also performed in the delayed-choice regime, using the procedure described above. We finally obtain independent measurement of $D$ and $V$ for different values of the reflection coefficient $R$, randomly applied to the interferometer.
The final results, depicted on Fig. 4, leads to $V^{2}+D^{2}=0.97 \pm 0.03$, close to the ideal balance between $V$ and $D$ constrained by inequality~\eqref{ComplIneq}, eventhough each quantity varies from $0$ to $1$.\\
\indent The effects observed in this delayed-choice experiment are in perfect agreement with quantum mechanics predictions. No change is observed between a so called ``normal-choice'' experiment and the ``delayed-choice'' version. It demonstrates that the complementarity principle cannot be interpreted in a naive way, assuming that the photon at the input of the interferometer could adjust its nature according to the experimental setup installed. As Bohr pointed out~\cite{{BohrConclusion}}, \textit{``it obviously can make no difference as regards observable effects obtainable by a definite experimental arrangement, whether our plans of constructing or handling the instrument are fixed beforehand or whether we prefer to postpone the completion of our planning until a later moment when the particle is already on its way from one instrument to another''}. Such intriguing property of quantum mechanics forces one to renounce to some common-sense representations of the physical rea\-lity.\\
\indent We warmly thank A. Clouqueur and A. Villing for the realization of the electronics, and J.-P. Madrange for all mechanical realization of the interferometer. This work is supported by Institut Universitaire de France.


\begin{references}
\bibitem{footnote no entanglement}
We stick to the original discussions on complementarity for single particles and do not consider schemes with pairs of entangled particles, where a measurements on one particle allows one to obtain which-path information on the other particle.
\bibitem{alpha}
The non-ideal value of the $\alpha$ parameter is due to residual background photoluminescence of the diamond sample and to its two-phonon Raman scattering line, which both produce uncorrelated photons associated to Poissonian statistics.
\end{references}
 \end{document}